\documentclass[a4paper,prb,twocolumn,groupedaddress,showpacs]{revtex4}

\usepackage[utf8]{inputenc}
\usepackage{amssymb}
\usepackage{amsmath}
\usepackage{color}
\usepackage{graphics, graphicx}
\usepackage{bbold}
\usepackage{psfrag}
\usepackage{mathcomp}
\usepackage{subfigure}
\usepackage{booktabs}
\usepackage{blindtext}
\usepackage{float}
\usepackage{import}
\usepackage{sansmath}

\usepackage[pdftitle={},
            pdfsubject={},
            pdfauthor={Jean Diehl},
            ]{hyperref}

\usepackage{color}
\usepackage{soul}

\newcommand{\ca}{CaFe$_2$As$_2$}

\begin{document}

\author{Jean Diehl}
\email[]{jdiehl@itp.uni-frankfurt.de}
\author{Steffen Backes}
\author{Daniel Guterding}
\author{Harald O. Jeschke}
\author{Roser Valent\'{\i}}
\affiliation{Institut f\"ur Theoretische Physik, Goethe-Universit\"at Frankfurt, Max-von-Laue-Str. 1, 60438 Frankfurt am Main, Germany}

\date{\today}

% 71.15.Mb Density functional theory, functionals etc.
% 71.18.+y Fermi surface: calculations and measurements; effective mass, gfactor
% 71.27.+a Strongly correlated electron systems; heavy fermions
% 74.70.Xa Pnictides and chalcogenides

\
\vspace{0.5cm}
\

\title{Correlation effects in the tetragonal and collapsed tetragonal phase of CaFe$_2$As$_2$}

\begin{abstract}
  We investigate the role of correlations in the tetragonal and
  collapsed tetragonal phases of {CaFe$_2$As$_2$} by performing charge
  self-consistent DFT+DMFT (density functional theory combined with
  dynamical mean-field theory) calculations. While the topology of the
  Fermi surface is basically unaffected by the inclusion of
  correlation effects, we find important orbital-dependent mass
  renormalizations which show good agreement with recent
  angle-resolved photoemission (ARPES) experiments. Moreover, we
  observe a markedly different behavior of these quantities between
  the low-pressure tetragonal and the high-pressure collapsed
  tetragonal phase.  We attribute these effects to the increased
  hybridization between the iron- and arsenic orbitals as one enters
  the collapsed tetragonal phase.
\end{abstract}

\maketitle

\section{INTRODUCTION}
\label{sec:introduction}
The iron-pnictide superconductor {\ca} belongs to the so-called 122
family, $A$Fe$_2$As$_2$ (e.g. $A= {\rm Ba}$, Sr, Ca) which
crystallizes in the $\mathrm{ThCr_2Si_2}$ structure with the
$I\,4/mmm$ space group.  {\ca} is tetragonal (TET) at room temperature
and ambient pressure and undergoes a structural phase transition to an
orthorhombic (ORT) phase upon cooling below
$170\,\mathrm{K}$.~\cite{Ronning08,Ni08a,Torikachvili08} Whereas the
tetragonal phase is non-magnetic, the orthorhombic phase shows a
stripe-like magnetic order~\cite{Diallo2009}. Upon application of
pressure, the appearance of a collapsed tetragonal (CT) phase
characterized by a collapse of the $c$ lattice parameter and a volume
shrinkage of about 5$\%$ with respect to the tetragonal phase was
observed.~\cite{Kreyssig08,Torikachvili08} First principles studies
have shown that for increasing pressures at low temperature the system
goes from the orthorhombic phase directly into the collapsed
tetragonal phase at $0.36\,\mathrm{GPa}$, whereas for higher
temperatures at high pressure the tetragonal phase is energetically
more favorable than the collapsed tetragonal phase.~\cite{Widom13}
Moreover, the ORT$\rightarrow$CT structural transition coincides with
the disappearance of the magnetic
moment~\cite{Yildirim2008,Zhang2009,Colonna11,Tomic12}.  Also in
BaFe$_2$As$_2$ such a collapsed tetragonal phase has been
theoretically~\cite{Colonna11,Tomic12,Backes13} predicted and
experimentally\cite{Uhoya10,Mittal11} observed, though at much higher
pressures of $\mathrm{27\,GPa}$~\cite{Mittal11} under hydrostatic
pressure conditions.

The appearance of a superconducting phase under pressure was reported
in {\ca} with a critical temperature of $10\,\mathrm{K}$ at
$0.69\,\mathrm{GPa}$~\cite{Park08}.  However, it was recently
established that the superconducting region is disjunct from the
non-magnetic collapsed tetragonal phase~\cite{Soh2013} and it is still
not entirely clear if superconductivity appears in the orthorhombic
phase or in a low temperature tetragonal phase that is stabilized by
special non-hydrostatic pressure conditions~\cite{Prokes2010}.  In
order to understand this behaviour, a lot of effort has been devoted
in the last years to investigate the electronic properties of the
collapsed tetragonal phase and its main differences compared to the
orthorhombic and tetragonal phases.  Angle-resolved photoemission
(ARPES) measurements for the orthorhombic and tetragonal phases in
{\ca} at ambient pressure were performed by Liu {\it et
  al.}~\cite{Liu09}, where a two- to three-dimensional transition in
the Fermi surface was observed, corresponding to the transition from
the tetragonal to the orthorhombic phase at low temperatures.
Measurements have also been performed for isostructural materials
which are in the collapsed tetragonal phase at ambient pressure:
$\mathrm{CaFe_2P_2}$~\cite{Coldea09} and
$\mathrm{Ca(Fe_{1-x}Rh_x)_2As_2}$~\cite{Tsubota13}. In both cases hole
pockets around the zone center $\Gamma$ present in the tetragonal
phase disappear in the collapsed tetragonal phase.

Only very recently {\ca} samples could be grown in the collapsed
tetragonal phase at ambient pressure by introducing internal
strain~\cite{Dhaka14}. In the same work, the authors performed
detailed ARPES measurements and found that collapsed tetragonal {\ca}
shows a similar behavior to $\mathrm{CaFe_2P_2}$ and
$\mathrm{Ca(Fe_{1-x}Rh_x)_2As_2}$, namely the disappearance of the
hole pockets at the $\Gamma$ point.  While density functional theory
(DFT) calculations correctly predict this
feature~\cite{Coldea09,Tsubota13,Tomic12}, ARPES measurements show a
strong band renormalization compared to the DFT calculations.

In order to investigate the origin of this discrepancy, we present in
this work an analysis of the electronic structure of tetragonal and
collapsed tetragonal phases of {\ca} by combining DFT in the GGA
approximation with dynamical mean-field theory (GGA+DMFT). This method
has been proven to provide a good description of correlation effects
in a few families of Fe-based
superconductors~\cite{Aichhorn10,Aichhorn11,Yin2011a,Ferber12b,Ferber2012a,Werner12}.
While the 122 family has been argued to be less correlated than the
so-called 111 or 11 families~\cite{Yin2011a}, we will show that also
in {\ca} correlations are necessary to understand the renormalization
of the bands, where we find a distinct change of orbital-dependent
mass enhancements in the transition from the tetragonal to the
collapsed tetragonal phase.

\begin{table}
  \caption{\label{tab:structures} Lattice parameters for the experimentally
    measured tetragonal and collapsed tetragonal
    structure from Ref.~\onlinecite{Kreyssig08}.}
    \begin{ruledtabular}
        \begin{tabular}{lccc}
         & ${\mathrm{TET}}$ & ${\mathrm{CT}}$  \\
         & $I\,4/mmm$  & $I\,4/mmm$\\
        \hline
        $T$ (K) & 250 & 50  \\
        $p$ (GPa) & 0.0 & 0.35 \\
        \hline
        $a$ ($b$) (\AA)  & 3.8915 & 3.9792  \\
        $c$ (\AA)  & 11.690  & 10.6073 \\
        $z_\text{As}$& 0.372 & 0.3663\\
        $V$  (\AA$^3$)  & 177.03  & 167.96\\
        \end{tabular}
    \end{ruledtabular}
\end{table}

\section{METHODS}
\label{sec:methods}

For our fully charge self-consistent GGA+DMFT calculations we consider
the tetragonal and collapsed tetragonal structures obtained by neutron
diffraction experiments~\cite{Kreyssig08}.  Lattice parameters and As
$z$ position are shown in Table~\ref{tab:structures}.

The DFT calculations were performed with the
\textsc{WIEN2k}~\cite{Blaha01} implementation of the full-potential
linear augmented plane wave (FLAPW) method. As exchange-correlation
functional we considered the generalized gradient
approximation~\cite{Perdew96} (GGA). The self-consistency cycle
employed 726 $k$-points in the irreducible Brillouin zone, resulting in
a $21\times 21\times 21$ $k$ mesh in the conventional Brillouin zone, and a
$R_{mt}k_{max} = 7.0$.  For the projection of the Bloch wave functions
to the localized Fe $3d$ orbitals we used our own implementation of
the method described in Ref.~\onlinecite{Aichhorn09,Ferber2014}.  The
energy window for the projection was chosen to be in the range from
$-5.9$ to $16.0\,\mathrm{eV}$ ($-6.3$ to $16.0\,\mathrm{eV}$) for the
tetragonal and collapsed tetragonal structures.  We were able to set
the lower energy boundary in a gap in the density of states (DOS).
The impurity problem was solved with a continuous-time quantum Monte
Carlo method in the hybridization expansion~\cite{Werner06} as
implemented in the ALPS~\cite{ALPS11,Gull11a} project.  Calculations
were done at $\beta = 40\,\mathrm{eV^{-1}}$ with $2\times 10^6$ Monte
Carlo sweeps.  For the double counting correction the fully localized
limit~\cite{Anisimov93,Dudarev98} (FLL) scheme was used, although the
around mean field~\cite{Czyzyk94} (AMF) scheme led to comparable
results with only slightly less renormalized masses.  The interaction
parameters are used in the definition of the Slater
integrals~\cite{Liechtenstein95} $F^k$ with $U=F^0$ and
$J=(F^2+F^4)/14$. For the onsite correlation we consider a value of
$U=4\,\mathrm{eV}$ and for Hund's rule coupling $J=0.8\,\mathrm{eV}$
and we analyze the dependency of our results on variations of these
parameters.  For the analytic continuation of the Monte Carlo data on
the imaginary time axis we used a combination of Pad\'e-approximation
and a fourth order polynomial fit to the first eight Matsubara
frequencies to obtain real frequency data.

\begin{figure}[hbp]
  \includegraphics[width=0.5\textwidth]{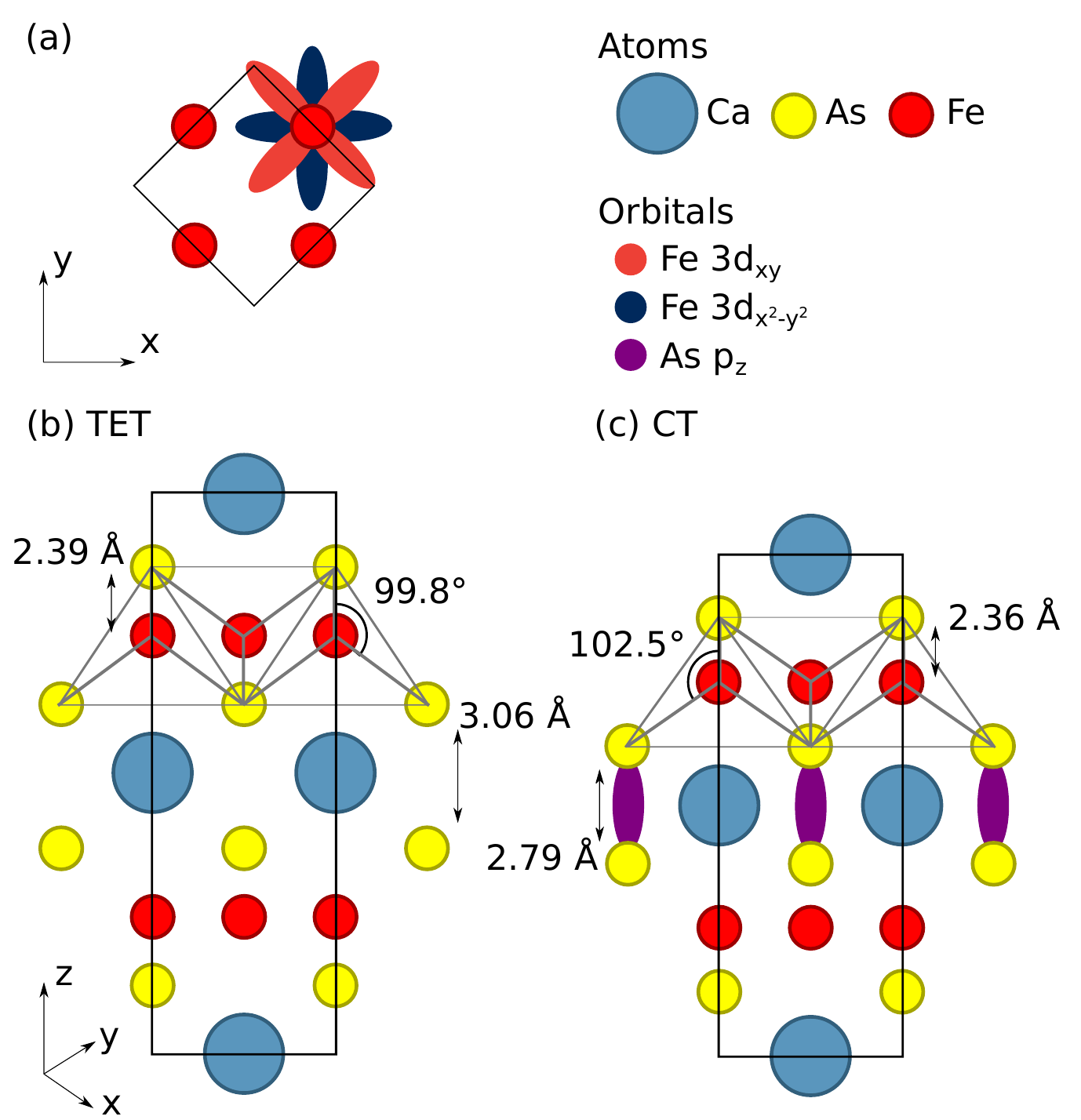}
  \caption{(Color online) Sketch of (a) the rotated coordinate system
    with the unit vectors pointing to neighboring iron atoms in the
    $xy$-plane. In this projection, Fe $3d_{x^2-y^2}$ orbitals point
    to neighboring Fe atoms, and Fe $3d_{xy}$ orbitals point towards
    As atoms. The side-views of the structures in the TET phase in (b)
    and the CT phase (c) show the collapse along the $z$-direction,
    allowing the As atoms to form As dimers in the CT phase when the
    As-As distance decreases.}
  \label{fig:orbitalssktech}
\end{figure}

In the projection of the Fe $3d$ orbital character, we use a
coordinate system which is rotated by $45^\circ$ around the $z$-axis
with respect to the conventional $I\,4/mmm$ unit cell so that $x$- and
$y$-axis point towards neighboring Fe atoms as shown in
Fig.~\ref{fig:orbitalssktech} (a).  In the band structure and Fermi
surface plots we choose the usual high symmetry points $X$, $M$ and
$Z$ of the $P\,4/nmm$ space group to facilitate comparison with the
other families of iron pnictides.

\section{RESULTS}
\label{sec:results}

\subsection{Band structure and spectral function}
\label{subsec:res:bandstructure}

\begin{figure}[t]
  \centering
  \includegraphics[width=0.48\textwidth, trim=0cm 0.5cm 0cm
  0cm]{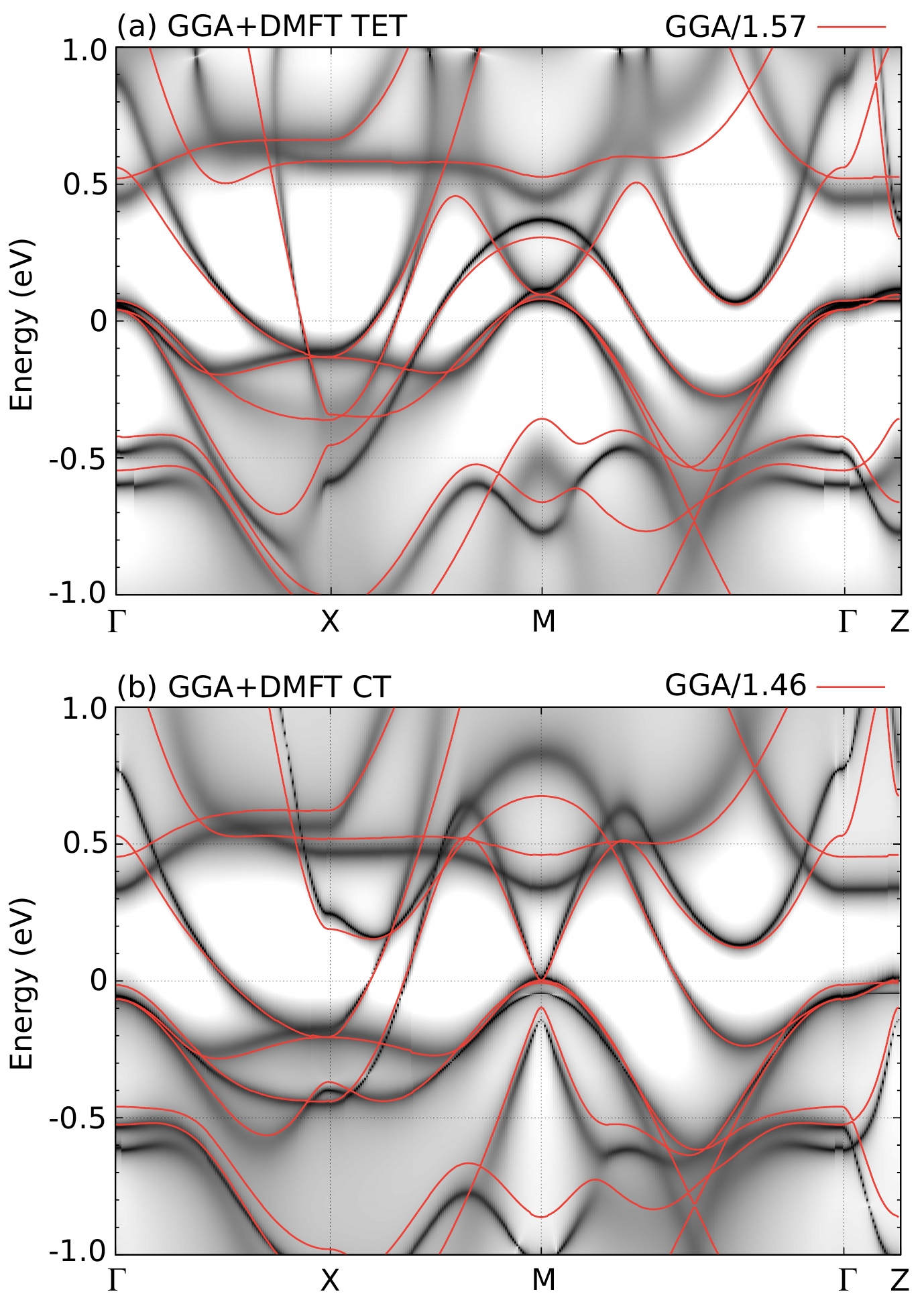}
  \caption{(Color online) Comparison between GGA band structure (red
    lines) divided by the orbitally averaged mass enhancements and the
    momentum resolved spectral function (greyscale in arbitrary units)
    from GGA+DMFT for (a) the experimental TET structure and (b) the CT
    structure.}
    \label{fig:bandstructure-gga}
\end{figure}

In Fig.~\ref{fig:bandstructure-gga} we show a comparison of the DFT
(GGA) band structure calculations and the spectral function obtained
with GGA+DMFT.  We find that correlations mostly renormalize bands in
both structures without introducing significant band shifts or
altering the topology of the Fermi surface.  In the tetragonal phase
we observe in both DFT (GGA) as well as GGA+DMFT calculations the
presence of three hole bands crossing the Fermi level at the zone
center $\Gamma$, two electron pockets at $X$ and three well-defined
hole pockets at the zone corner $M$ formed by strongly dispersive hole
bands with a large outer pocket and two smaller inner pockets almost
identical in size.  In the collapsed tetragonal phase the bands at
$\Gamma$ are pushed below the Fermi level in agreement with
experiments~\cite{Coldea09,Tomic12,Tsubota13,Dhaka14}, the inner
electron pocket at $X$ is pushed up to positive energies leaving only
the slightly enlarged outer electron pocket present.  At $M$ the bands
forming the inner two hole pockets are pushed onto the Fermi level,
leaving two extremely shallow bands of which only one just barely
crosses $E_F$.  GGA+DMFT introduces a significant separation between
the two bands not observed in the DFT(GGA) calculations.  This is a
result of the orbital dependent correlations introduced by DMFT.

\begin{figure}[t]
  \centering
  \includegraphics[width=0.48\textwidth, trim=0cm 0.5cm 0cm
  0cm]{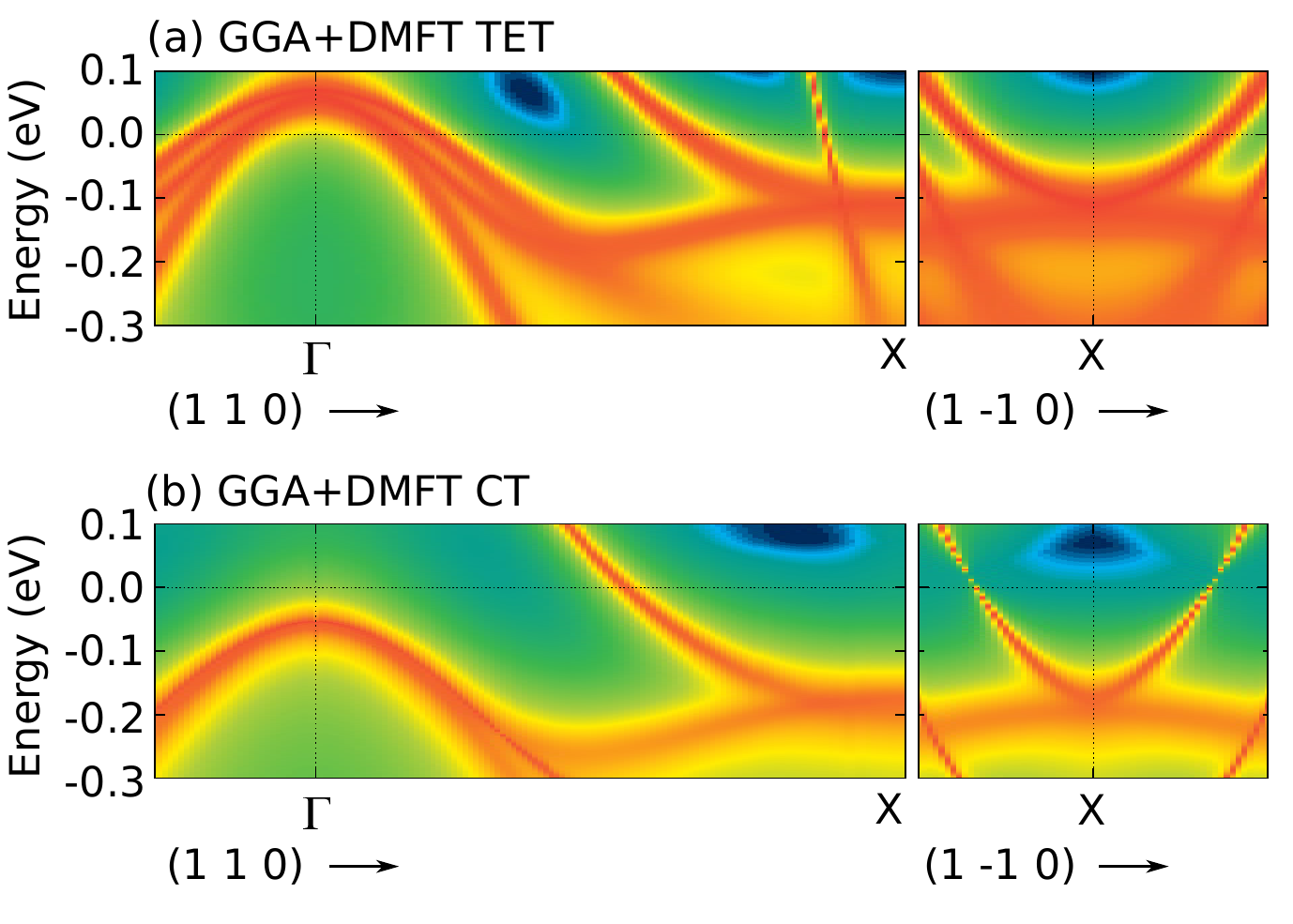}
  \caption{(Color online) Momentum resolved spectral function around
    the Fermi level along the same path in the Brillouin zone as in
    Ref.~\onlinecite{Dhaka14} for (a) the experimental TET structure and
    (b) the CT structure.}
\label{fig:bandstructure}
\end{figure}

In Fig. \ref{fig:bandstructure} we show the GGA+DMFT results for the
same energy range and along the same path in the Brillouin zone as in
Ref.~\onlinecite{Dhaka14} in order to allow a better comparison to the
ARPES measurements. We find a good agreement between ARPES and our
GGA+DMFT calculation in both the tetragonal and the collapsed
tetragonal phases albeit GGA+DMFT finds a smaller band renormalization
than the value extracted from ARPES. Our band renormalizations are
about a factor of 1.7 compared to GGA masses while the ARPES
measurements report a factor of 5. This suggests that other possible
contributions not considered in DMFT may also be important for the
description of the electronic behavior of {\ca} like non-local
correlations and electron-phonon interactions.

\begin{figure}[htbp]
\includegraphics[width=0.5\textwidth]{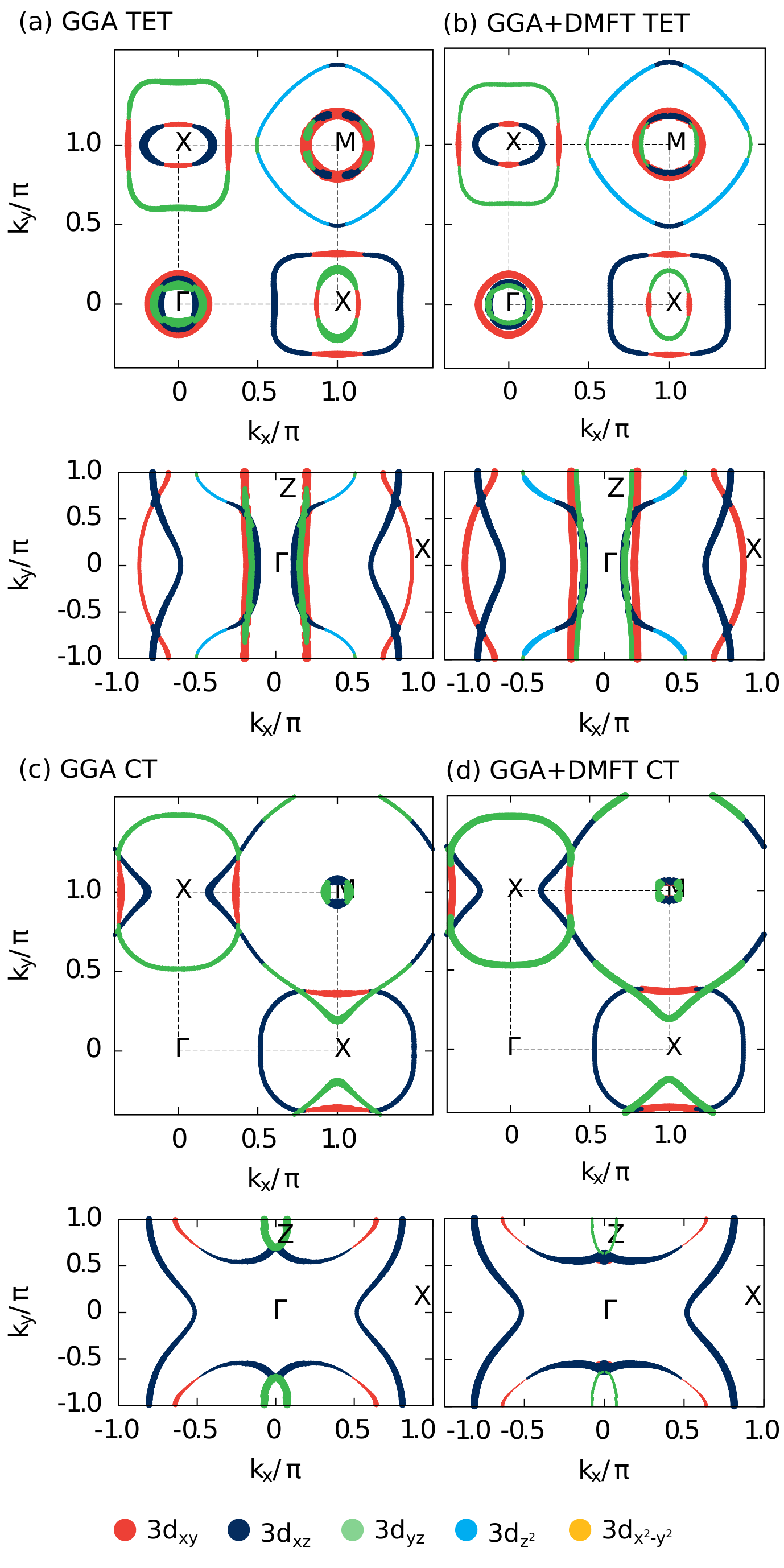}
\caption{(Color online) Comparison of the Fermi surface from GGA
  (left) and GGA+DMFT (right) along a plane at $k_z=0$ and a vertical
  cut through the $\Gamma$ and $\mathrm{X}$ point.}
    \label{fig:fermi}
\end{figure}

The GGA+DMFT Fermi surface for {\ca} shows only slight changes
compared to DFT(GGA) (see Fig.~\ref{fig:fermi}) and agrees reasonably
well with ARPES measurements~\cite{Dhaka14}.  The main features of the
collapsed tetragonal phase are the disappearance of the hole pockets
at $\Gamma$ as well as a change from a more two-dimensional shape in
the tetragonal phase to a three-dimensional shape in the collapsed
tetragonal phase (compare the cuts along a plane parallel to the $z$
direction in Figs.~\ref{fig:fermi} (a) and (b)) due to increasing Fe
3$d$-As 4$p$ hybridizations.

\subsection{Mass enhancements and sensitivity to interaction parameters}
\label{subsec:res:massenhancements}

We calculate the effective masses directly from the impurity
self-energy via
\begin{equation}
    \frac{m^\ast}{m_\mathrm{GGA}} = 1 - \left.\frac{\partial\mathrm{Im}\Sigma(i\omega)}{\partial \omega}\right|_{\omega\rightarrow 0^+}.
    \label{eq:meff}
\end{equation}

For the interaction parameters set to $U=4\,\mathrm{eV}$ and
$J=0.8\,\mathrm{eV}$ we obtain mass renormalizations between $1.2$ and
$1.7$ as shown in Tab. \ref{tab:meff} for the different orbital
characters. Mass renormalizations are strongest for the $t_{2g}$
orbitals Fe 3$d_{xy}$ and 3$d_{xz/yz}$ in the tetragonal phase while
the $e_g$ orbitals 3$d_{z^2}$ and 3$d_{x^2-y^2}$ are less renormalized
both in the tetragonal and collapsed tetragonal phases.

\begin{figure}[t]
  \centering
  \includegraphics[width=0.48\textwidth, trim=0cm 2cm 2cm 0cm]{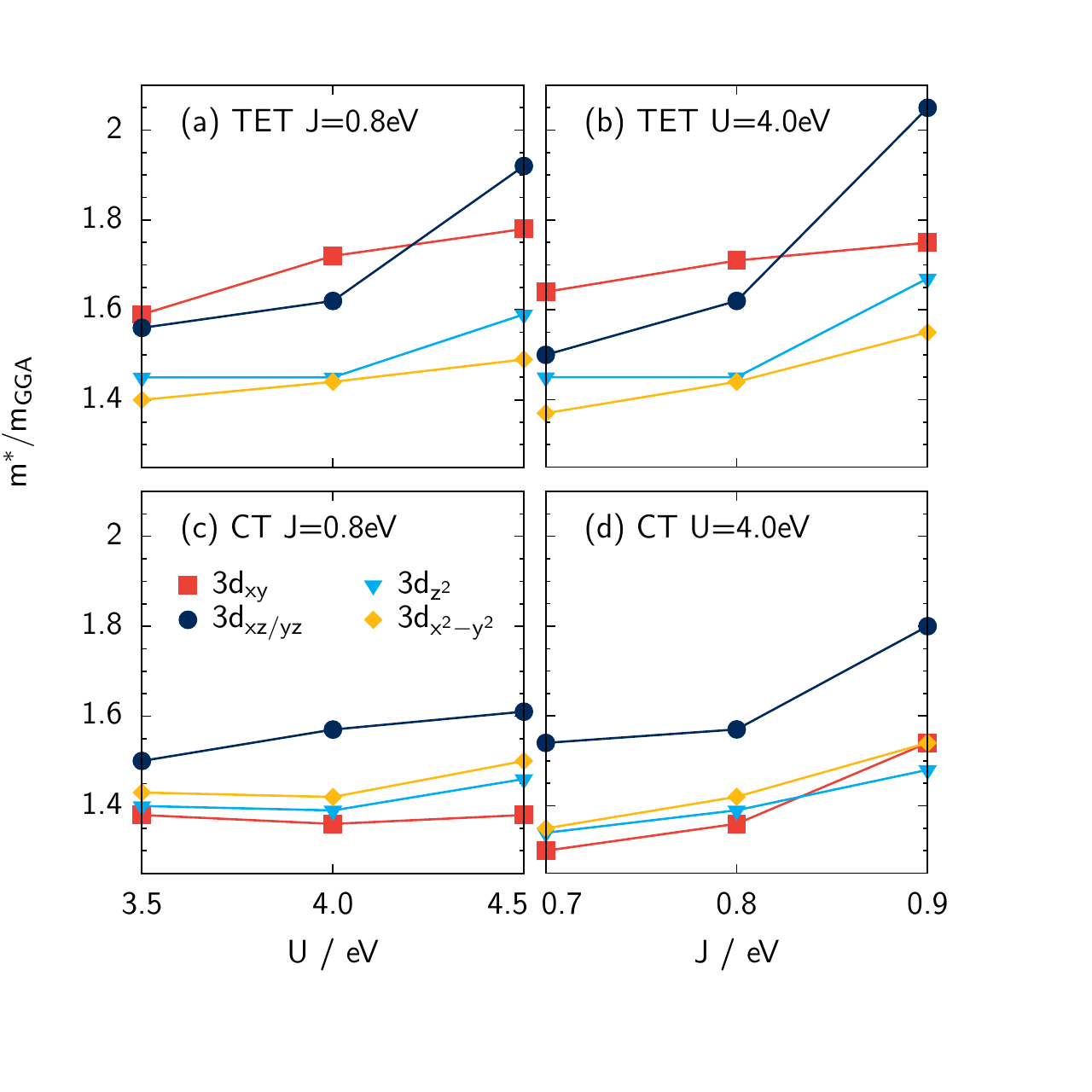}
  \caption{(Color online) Sensitivity of effective masses
    $\mathrm{m^\ast / m_{GGA}}$ with respect to changes in the
    interaction parameters. (a), (c) show variations in $U$ in the
    tetragonal and the collapsed tetragonal phases respectively, and
    (b), (d) show variations in $J$.}
  \label{fig:meffcomp}
\end{figure}

As shown in Table~\ref{tab:meff} and Fig.~\ref{fig:meffcomp} we
observe a change in the strengths of the mass renormalizations.
Interestingly, the iron Fe 3$d_{xy}$ orbital undergoes a change from
being the most strongly renormalized orbital in the tetragonal phase
to the least renormalized orbital in the collapsed tetragonal phase.
This can be understood in terms of increased hybridization in the
collapsed tetragonal phase.  The structural collapse in this phase is
assisted by a formation of As 4$p_z$-As 4$p_z$ bonds~\cite{Tomic12}
between the Fe-As layers as shown in Fig.~\ref{fig:orbitalssktech} (b)
and (c) with a strong bonding-antibonding splitting of the As 4$p_z$
bands.  The Fe 3$d_{xy}$ orbitals, which are pointing in the direction
of the As atoms, become less localized in the collapsed tetragonal
phase due to increased Fe 3$d_{xy}$-Fe 3$d_{xy}$ as well as Fe
3$d_{xy}$-As 4$p_x$ and 4$p_y$ hybridizations~\cite{milandiscuss}.  This higher degree of
delocalization leads to less mass renormalization upon inclusion of
correlations.

\begin{table}[t]
  \caption{\label{tab:meff} Mass renormalizations calculated with GGA+DMFT
    for the Fe $3d$ orbitals.}
  \begin{ruledtabular}
    \begin{tabular}{lcccc}
    & $\mathrm{d_{z^2}}$ & $\mathrm{d_{x^2-y^2}}$ & $\mathrm{d_{xy}}$ & $\mathrm{d_{xz/yz}}$ \\
    \hline
    Tetragonal & 1.45 & 1.44 & 1.72 & 1.62 \\
    \hline
    Collapsed tetragonal & 1.39 & 1.42 & 1.36 & 1.57 \\
    %\hline
    %$\mathrm{CT_{DFT}}$ & 1.32 & 1.33 & 1.29 & 1.34 \\
    \end{tabular}
  \end{ruledtabular}
\end{table}

By varying the interaction parameters $U$ and $J$ we have investigated
their influence on the effective masses. The effective masses show
stronger dependencies on the Hund's rule coupling $J$ than the Hubbard
$U$ as shown in Fig. \ref{fig:meffcomp} and as already reported for
other members of the iron pnictides~\cite{Ferber12b,Haule09}. Our
results are stable for all values of the chosen interaction parameters
and, except for the stronger band renormalization, we observe only
very small qualitative changes in the Fermi surface.

\begin{figure}[t]
    \begin{center}
    \includegraphics[width=0.48\textwidth, trim=0cm 1.5cm 0.5cm 0cm]{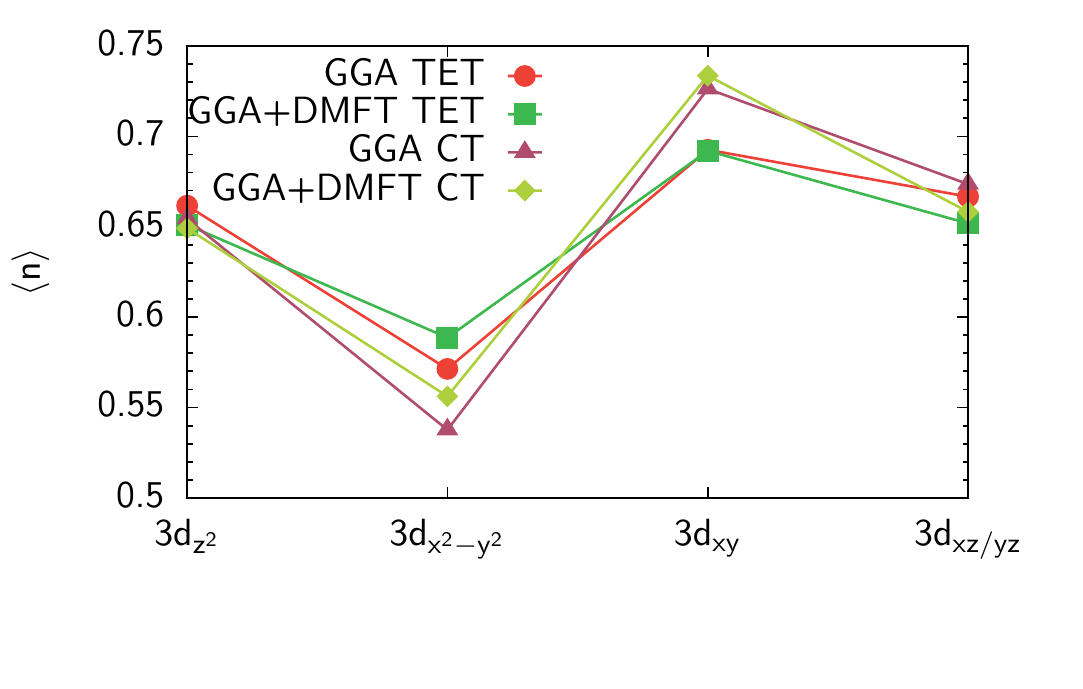}
    \end{center}
    \caption{(Color online) Orbital resolved occupation numbers for the
    GGA and the GGA+DMFT calculation.}
    \label{fig:occ}
\end{figure}

In Fig. \ref{fig:occ} we show a comparison of the occupation numbers
between the GGA and the GGA+DMFT results for the tetragonal and the
collapsed tetragonal phases.  The 3$d_{xy}$ and 3$d_{xz/yz}$ show the
largest occupation with respect to 3$d_{z^2}$ and 3$d_{x^2-y^2}$
reflecting the crystal field splitting in $t_{2g}$ and $e_g$ orbitals.
At the GGA level the transition from tetragonal to collapsed
tetragonal phase implies a pronounced increase of charge occupation of
the 3$d_{xy}$ orbital and to a lesser extent of the 3$d_{xz/yz}$,
while the occupation for the $e_g$ states decreases. This can also be
understood in terms of the change in hybridizations as explained
above, where due to the enhanced delocalization of the 3$d_{xy}$
electrons in the collapsed tetragonal phase the 3$d_{xy}$ orbital
becomes less correlated.  Regarding the GGA versus GGA+DMFT
occupations we observe only little changes and a general trend of
electronic charge being shifted from the most correlated orbitals to
the less correlated orbitals, as expected, with the total charge on
the Fe $3d$ orbitals staying basically identical to the DFT
calculation.

Recently, we became aware of the ARPES investigations by Gofryk {\it
  et al.}~\cite{Gofryk14}, who reported a distinct increase of the
effective masses of the bands around the $\Gamma$-point when entering
the collapsed tetragonal phase.  In order to understand this, we
calculated the effective masses $m_{GGA}/m_e$ of the three hole bands
around the $\Gamma$-point according to the method we described in a
previous article~\cite{k122}. In the tetragonal phase we obtained
1.11, 1.62 and 1.71$m_e$, while in the CT phase we obtained 1.53,
2.00, 2.89$m_e$, with the bands ordered from higher to lower binding
energies. Thus, already at the GGA level the trend of increasing
renormalization of the bands around $\Gamma$ in the CT phase is
correctly described, albeit the absolute values are lower compared to
what was reported from experiment~\cite{Gofryk14}. Therefore, we
conclude that the observed increase in band renormalizations from the
tetragonal to the CT phase around $\Gamma$ is mostly due to stronger
hybridizations in the collapsed tetragonal phase, as discussed in this
section, leading to a shift of the hole bands below the Fermi level.
Electronic correlations contribute further only to a minor degree to
the effective electronic mass of the bands around $\Gamma$, which we
attribute to the fact that {\ca} is a weakly to moderately correlated
metal.

\section{CONCLUSIONS}
\label{sec:conclusion}
We have performed charge self-consistent GGA+DMFT calculations for
{\ca} in the tetragonal and collapsed tetragonal phases.  We observe
that while the topology of the Fermi surface in both phases remains
nearly unaffected, the orbital-selective mass renormalizations of a
factor $1.3$ to $1.7$ introduced by GGA+DMFT improve the agreement of
the calculations with ARPES experiments.  The analysis of the
influence of the tetragonal to collapsed tetragonal transition on the
orbital-dependent effective masses shows that Fe $3d_{xy}$ changes
from being the most strongly correlated orbital in the tetragonal
phase to being the least correlated one in the collapsed tetragonal
phase. We attribute this to the change in hybridization of the Fe $3d$
orbitals in the collapsed tetragonal phase, where due to the decreased
distance of the Fe-As layers the hybridization for the Fe 3$d_{xy}$-Fe
3$d_{xy}$ as well as Fe 3$d_{xy}$-As 4$p_x$ and 4$p_y$ orbitals
increases, rendering the Fe 3$d_{xy}$ less localized and thus less
correlated.  The orbital occupations confirm this trend and show a
higher occupation for the Fe 3$d_{xy}$ orbital in the collapsed
tetragonal phase.

With these observations we conclude that correlation effects beyond
DFT(GGA) as introduced by GGA+DMFT are needed even for weakly correlated
pnictides like {\ca} in order to understand the orbital-selective mass
renormalizations observed in ARPES. However, we also observe that such
a description is, nevertheless, still insufficient for explaining
the large mass renormalizations observed experimentally.  We attribute
this discrepancy to possible non-local correlations as well as
phononic effects and this will be a subject of future investigations.

During finalization of this manuscript we became aware of another
preprint of a DFT+DMFT study of {\ca}~\cite{mandal14}, where the
authors also find the same trend of reduced renormalization in the CT
phase and their results agree, except for minor quantitative
differences, with our findings.

\acknowledgments 
%\begin{center}\emph{Acknowledgments}\end{center}

The authors gratefully acknowledge Milan Tomi{\'c}, Paul Canfield and 
Peter Hirschfeld for helpful discussions and
the Deutsche Forschungsgemeinschaft for financial support through
grant SPP 1458.  We thank the centre for scientific computing (CSC,
LOEWE-CSC) in Frankfurt for computing facilities.

%\bibliography{library}

\begin{thebibliography}{37}
\expandafter\ifx\csname natexlab\endcsname\relax\def\natexlab#1{#1}\fi
\expandafter\ifx\csname bibnamefont\endcsname\relax
  \def\bibnamefont#1{#1}\fi
\expandafter\ifx\csname bibfnamefont\endcsname\relax
  \def\bibfnamefont#1{#1}\fi
\expandafter\ifx\csname citenamefont\endcsname\relax
  \def\citenamefont#1{#1}\fi
\expandafter\ifx\csname url\endcsname\relax
  \def\url#1{\texttt{#1}}\fi
\expandafter\ifx\csname urlprefix\endcsname\relax\def\urlprefix{URL }\fi
\providecommand{\bibinfo}[2]{#2}
\providecommand{\eprint}[2][]{\url{#2}}

\bibitem[{\citenamefont{Ronning et~al.}(2008)\citenamefont{Ronning, Klimczuk,
  Bauer, Volz, and Thompson}}]{Ronning08}
\bibinfo{author}{\bibfnamefont{F.}~\bibnamefont{Ronning}},
  \bibinfo{author}{\bibfnamefont{T.}~\bibnamefont{Klimczuk}},
  \bibinfo{author}{\bibfnamefont{E.~D.} \bibnamefont{Bauer}},
  \bibinfo{author}{\bibfnamefont{H.}~\bibnamefont{Volz}}, \bibnamefont{and}
  \bibinfo{author}{\bibfnamefont{J.~D.} \bibnamefont{Thompson}},
  \bibinfo{journal}{J. Phys. Condens. Matter} \textbf{\bibinfo{volume}{20}},
  \bibinfo{pages}{322201} (\bibinfo{year}{2008}).

\bibitem[{\citenamefont{Ni et~al.}(2008)\citenamefont{Ni, Nandi, Kreyssig,
  Goldman, Mun, Bud'ko, and Canfield}}]{Ni08a}
\bibinfo{author}{\bibfnamefont{N.}~\bibnamefont{Ni}},
  \bibinfo{author}{\bibfnamefont{S.}~\bibnamefont{Nandi}},
  \bibinfo{author}{\bibfnamefont{A.}~\bibnamefont{Kreyssig}},
  \bibinfo{author}{\bibfnamefont{A.~I.} \bibnamefont{Goldman}},
  \bibinfo{author}{\bibfnamefont{E.~D.}~\bibnamefont{Mun}},
  \bibinfo{author}{\bibfnamefont{S.~L.}~\bibnamefont{Bud'ko}}, \bibnamefont{and}
  \bibinfo{author}{\bibfnamefont{P.~C.}~\bibnamefont{Canfield}},
  \bibinfo{journal}{Phys. Rev. B} \textbf{\bibinfo{volume}{78}},
  \bibinfo{pages}{014523} (\bibinfo{year}{2008}).

\bibitem[{\citenamefont{Torikachvili et~al.}(2008)\citenamefont{Torikachvili,
  Bud'ko, Ni, and Canfield}}]{Torikachvili08}
\bibinfo{author}{\bibfnamefont{M.~S.} \bibnamefont{Torikachvili}},
  \bibinfo{author}{\bibfnamefont{S.~L.} \bibnamefont{Bud'ko}},
  \bibinfo{author}{\bibfnamefont{N.}~\bibnamefont{Ni}}, \bibnamefont{and}
  \bibinfo{author}{\bibfnamefont{P.~C.} \bibnamefont{Canfield}},
  \bibinfo{journal}{Phys. Rev. Lett.} \textbf{\bibinfo{volume}{101}},
  \bibinfo{pages}{057006} (\bibinfo{year}{2008}).

\bibitem[{\citenamefont{Diallo et~al.}(2009)\citenamefont{Diallo, Antropov,
  Perring, Broholm, Pulikkotil, Ni, Bud'ko, Canfield, Kreyssig, Goldman
  et~al.}}]{Diallo2009}
\bibinfo{author}{\bibfnamefont{S.~O.} \bibnamefont{Diallo}},
  \bibinfo{author}{\bibfnamefont{V.~P.} \bibnamefont{Antropov}},
  \bibinfo{author}{\bibfnamefont{T.~G.} \bibnamefont{Perring}},
  \bibinfo{author}{\bibfnamefont{C.}~\bibnamefont{Broholm}},
  \bibinfo{author}{\bibfnamefont{J.~J.} \bibnamefont{Pulikkotil}},
  \bibinfo{author}{\bibfnamefont{N.}~\bibnamefont{Ni}},
  \bibinfo{author}{\bibfnamefont{S.~L.} \bibnamefont{Bud'ko}},
  \bibinfo{author}{\bibfnamefont{P.~C.} \bibnamefont{Canfield}},
  \bibinfo{author}{\bibfnamefont{A.}~\bibnamefont{Kreyssig}},
  \bibinfo{author}{\bibfnamefont{A.~I.} \bibnamefont{Goldman}},
  \bibnamefont{et~al.}, \bibinfo{journal}{Phys. Rev. Lett.}
  \textbf{\bibinfo{volume}{102}}, \bibinfo{pages}{187206}
  (\bibinfo{year}{2009}).

\bibitem[{\citenamefont{Kreyssig et~al.}(2008)\citenamefont{Kreyssig, Green,
  Lee, Samolyuk, Zajdel, Lynn, Bud'ko, Torikachvili, Ni, Nandi
  et~al.}}]{Kreyssig08}
\bibinfo{author}{\bibfnamefont{A.}~\bibnamefont{Kreyssig}},
  \bibinfo{author}{\bibfnamefont{M.}~\bibnamefont{Green}},
  \bibinfo{author}{\bibfnamefont{Y.}~\bibnamefont{Lee}},
  \bibinfo{author}{\bibfnamefont{G.}~\bibnamefont{Samolyuk}},
  \bibinfo{author}{\bibfnamefont{P.}~\bibnamefont{Zajdel}},
  \bibinfo{author}{\bibfnamefont{J.}~\bibnamefont{Lynn}},
  \bibinfo{author}{\bibfnamefont{S.}~\bibnamefont{Bud'ko}},
  \bibinfo{author}{\bibfnamefont{M.}~\bibnamefont{Torikachvili}},
  \bibinfo{author}{\bibfnamefont{N.}~\bibnamefont{Ni}},
  \bibinfo{author}{\bibfnamefont{S.}~\bibnamefont{Nandi}},
  \bibnamefont{et~al.}, \bibinfo{journal}{Phys. Rev. B}
  \textbf{\bibinfo{volume}{78}}, \bibinfo{pages}{184517}
  (\bibinfo{year}{2008}).

\bibitem[{\citenamefont{Widom and Quader}(2013)}]{Widom13}
\bibinfo{author}{\bibfnamefont{M.}~\bibnamefont{Widom}} \bibnamefont{and}
  \bibinfo{author}{\bibfnamefont{K.}~\bibnamefont{Quader}},
  \bibinfo{journal}{Phys. Rev. B} \textbf{\bibinfo{volume}{88}},
  \bibinfo{pages}{045117} (\bibinfo{year}{2013}).

\bibitem[{\citenamefont{Yildirim}(2008)}]{Yildirim2008}
\bibinfo{author}{\bibfnamefont{T.}~\bibnamefont{Yildirim}},
  \bibinfo{journal}{Phys. Rev. Lett.} \textbf{\bibinfo{volume}{101}},
  \bibinfo{pages}{057010} (\bibinfo{year}{2008}).

\bibitem[{\citenamefont{Zhang et~al.}(2009)\citenamefont{Zhang, Kandpal,
  Opahle, Jeschke, and Valent\'\i}}]{Zhang2009}
\bibinfo{author}{\bibfnamefont{Y.-Z.} \bibnamefont{Zhang}},
  \bibinfo{author}{\bibfnamefont{H.~C.} \bibnamefont{Kandpal}},
  \bibinfo{author}{\bibfnamefont{I.}~\bibnamefont{Opahle}},
  \bibinfo{author}{\bibfnamefont{H.~O.} \bibnamefont{Jeschke}},
  \bibnamefont{and}
  \bibinfo{author}{\bibfnamefont{R.}\bibnamefont{Valenti}},
  \bibinfo{journal}{Phys. Rev. B} \textbf{\bibinfo{volume}{80}},
  \bibinfo{pages}{094530} (\bibinfo{year}{2009}).

\bibitem[{\citenamefont{Colonna et~al.}(2011)\citenamefont{Colonna, Profeta,
  Continenza, and Massidda}}]{Colonna11}
\bibinfo{author}{\bibfnamefont{N.}~\bibnamefont{Colonna}},
  \bibinfo{author}{\bibfnamefont{G.}~\bibnamefont{Profeta}},
  \bibinfo{author}{\bibfnamefont{A.}~\bibnamefont{Continenza}},
  \bibnamefont{and} \bibinfo{author}{\bibfnamefont{S.}~\bibnamefont{Massidda}},
  \bibinfo{journal}{Phys. Rev. B} \textbf{\bibinfo{volume}{83}},
  \bibinfo{pages}{094529} (\bibinfo{year}{2011}).

\bibitem[{\citenamefont{Tomic et~al.}(2012)\citenamefont{Tomic, Valenti, and
  Jeschke}}]{Tomic12}
\bibinfo{author}{\bibfnamefont{M.}~\bibnamefont{Tomic}},
  \bibinfo{author}{\bibfnamefont{R.}~\bibnamefont{Valenti}}, \bibnamefont{and}
  \bibinfo{author}{\bibfnamefont{H.~O.} \bibnamefont{Jeschke}},
  \bibinfo{journal}{Phys. Rev. B} \textbf{\bibinfo{volume}{85}},
  \bibinfo{pages}{094105} (\bibinfo{year}{2012}).

\bibitem[{\citenamefont{Backes and Jeschke}(2013)}]{Backes13}
\bibinfo{author}{\bibfnamefont{S.}~\bibnamefont{Backes}} \bibnamefont{and}
  \bibinfo{author}{\bibfnamefont{H.~O.} \bibnamefont{Jeschke}},
  \bibinfo{journal}{Phys. Rev. B} \textbf{\bibinfo{volume}{88}},
  \bibinfo{pages}{075111} (\bibinfo{year}{2013}).

\bibitem[{\citenamefont{Uhoya et~al.}(2010)\citenamefont{Uhoya, Stemshorn,
  Tsoi, Vohra, Sefat, Sales, Hope, and Weir}}]{Uhoya10}
\bibinfo{author}{\bibfnamefont{W.}~\bibnamefont{Uhoya}},
  \bibinfo{author}{\bibfnamefont{A.}~\bibnamefont{Stemshorn}},
  \bibinfo{author}{\bibfnamefont{G.}~\bibnamefont{Tsoi}},
  \bibinfo{author}{\bibfnamefont{Y.~K.} \bibnamefont{Vohra}},
  \bibinfo{author}{\bibfnamefont{A.~S.} \bibnamefont{Sefat}},
  \bibinfo{author}{\bibfnamefont{B.~C.} \bibnamefont{Sales}},
  \bibinfo{author}{\bibfnamefont{K.~M.} \bibnamefont{Hope}}, \bibnamefont{and}
  \bibinfo{author}{\bibfnamefont{S.~T.} \bibnamefont{Weir}},
  \bibinfo{journal}{Phys. Rev. B} \textbf{\bibinfo{volume}{82}},
  \bibinfo{pages}{144118} (\bibinfo{year}{2010}).

\bibitem[{\citenamefont{Mittal et~al.}(2011)\citenamefont{Mittal, Mishra,
  Chaplot, Ovsyannikov, Greenberg, Trots, Dubrovinsky, Su, Brueckel, Matsuishi
  et~al.}}]{Mittal11}
\bibinfo{author}{\bibfnamefont{R.}~\bibnamefont{Mittal}},
  \bibinfo{author}{\bibfnamefont{S.~K.} \bibnamefont{Mishra}},
  \bibinfo{author}{\bibfnamefont{S.~L.} \bibnamefont{Chaplot}},
  \bibinfo{author}{\bibfnamefont{S.~V.} \bibnamefont{Ovsyannikov}},
  \bibinfo{author}{\bibfnamefont{E.}~\bibnamefont{Greenberg}},
  \bibinfo{author}{\bibfnamefont{D.~M.} \bibnamefont{Trots}},
  \bibinfo{author}{\bibfnamefont{L.}~\bibnamefont{Dubrovinsky}},
  \bibinfo{author}{\bibfnamefont{Y.}~\bibnamefont{Su}},
  \bibinfo{author}{\bibfnamefont{T.}~\bibnamefont{Brueckel}},
  \bibinfo{author}{\bibfnamefont{S.}~\bibnamefont{Matsuishi}},
  \bibnamefont{et~al.}, \bibinfo{journal}{Phys. Rev. B}
  \textbf{\bibinfo{volume}{83}}, \bibinfo{pages}{054503}
  (\bibinfo{year}{2011}).

\bibitem[{\citenamefont{Park et~al.}(2008)\citenamefont{Park, Park, Lee,
  Klimczuk, Bauer, Ronning, and Thompson}}]{Park08}
\bibinfo{author}{\bibfnamefont{T.}~\bibnamefont{Park}},
  \bibinfo{author}{\bibfnamefont{E.}~\bibnamefont{Park}},
  \bibinfo{author}{\bibfnamefont{H.}~\bibnamefont{Lee}},
  \bibinfo{author}{\bibfnamefont{T.}~\bibnamefont{Klimczuk}},
  \bibinfo{author}{\bibfnamefont{E.~D.} \bibnamefont{Bauer}},
  \bibinfo{author}{\bibfnamefont{F.}~\bibnamefont{Ronning}}, \bibnamefont{and}
  \bibinfo{author}{\bibfnamefont{J.~D.} \bibnamefont{Thompson}},
  \bibinfo{journal}{J. Phys. Condens. Matter} \textbf{\bibinfo{volume}{20}},
  \bibinfo{pages}{322204} (\bibinfo{year}{2008}).

\bibitem[{\citenamefont{Soh et~al.}(2013)\citenamefont{Soh, Tucker, Pratt,
  Abernathy, Stone, Ran, Bud'ko, Canfield, Kreyssig, McQueeney
  et~al.}}]{Soh2013}
\bibinfo{author}{\bibfnamefont{J.~H.} \bibnamefont{Soh}},
  \bibinfo{author}{\bibfnamefont{G.~S.} \bibnamefont{Tucker}},
  \bibinfo{author}{\bibfnamefont{D.~K.} \bibnamefont{Pratt}},
  \bibinfo{author}{\bibfnamefont{D.~L.} \bibnamefont{Abernathy}},
  \bibinfo{author}{\bibfnamefont{M.~B.} \bibnamefont{Stone}},
  \bibinfo{author}{\bibfnamefont{S.}~\bibnamefont{Ran}},
  \bibinfo{author}{\bibfnamefont{S.~L.} \bibnamefont{Bud'ko}},
  \bibinfo{author}{\bibfnamefont{P.~C.} \bibnamefont{Canfield}},
  \bibinfo{author}{\bibfnamefont{A.}~\bibnamefont{Kreyssig}},
  \bibinfo{author}{\bibfnamefont{R.~J.} \bibnamefont{McQueeney}},
  \bibnamefont{et~al.}, \bibinfo{journal}{Phys. Rev. Lett.}
  \textbf{\bibinfo{volume}{111}}, \bibinfo{pages}{227002}
  (\bibinfo{year}{2013}).

\bibitem[{\citenamefont{Proke\ifmmode~\check{s}\else \v{s}\fi{}
  et~al.}(2010)\citenamefont{Proke\ifmmode~\check{s}\else \v{s}\fi{}, Kreyssig,
  Ouladdiaf, Pratt, Ni, Bud'ko, Canfield, McQueeney, Argyriou, and
  Goldman}}]{Prokes2010}
\bibinfo{author}{\bibfnamefont{K.}~\bibnamefont{Proke\v{s}}}, \bibinfo{author}{\bibfnamefont{A.}~\bibnamefont{Kreyssig}},
  \bibinfo{author}{\bibfnamefont{B.}~\bibnamefont{Ouladdiaf}},
  \bibinfo{author}{\bibfnamefont{D.~K.} \bibnamefont{Pratt}},
  \bibinfo{author}{\bibfnamefont{N.}~\bibnamefont{Ni}},
  \bibinfo{author}{\bibfnamefont{S.~L.} \bibnamefont{Bud'ko}},
  \bibinfo{author}{\bibfnamefont{P.~C.} \bibnamefont{Canfield}},
  \bibinfo{author}{\bibfnamefont{R.~J.} \bibnamefont{McQueeney}},
  \bibinfo{author}{\bibfnamefont{D.~N.} \bibnamefont{Argyriou}},
  \bibnamefont{and} \bibinfo{author}{\bibfnamefont{A.~I.}
  \bibnamefont{Goldman}}, \bibinfo{journal}{Phys. Rev. B}
  \textbf{\bibinfo{volume}{81}}, \bibinfo{pages}{180506}
  (\bibinfo{year}{2010}).

\bibitem[{\citenamefont{Liu et~al.}(2009)\citenamefont{Liu, Kondo, Ni,
  Palczewski, Bostwick, Samolyuk, Khasanov, Shi, Rotenberg, Bud'ko
  et~al.}}]{Liu09}
\bibinfo{author}{\bibfnamefont{C.}~\bibnamefont{Liu}},
  \bibinfo{author}{\bibfnamefont{T.}~\bibnamefont{Kondo}},
  \bibinfo{author}{\bibfnamefont{N.}~\bibnamefont{Ni}},
  \bibinfo{author}{\bibfnamefont{a.~A.} \bibnamefont{Palczewski}},
  \bibinfo{author}{\bibfnamefont{A.}~\bibnamefont{Bostwick}},
  \bibinfo{author}{\bibfnamefont{G.}~\bibnamefont{Samolyuk}},
  \bibinfo{author}{\bibfnamefont{R.}~\bibnamefont{Khasanov}},
  \bibinfo{author}{\bibfnamefont{M.}~\bibnamefont{Shi}},
  \bibinfo{author}{\bibfnamefont{E.}~\bibnamefont{Rotenberg}},
  \bibinfo{author}{\bibfnamefont{S.}~\bibnamefont{Bud'ko}},
  \bibnamefont{et~al.}, \bibinfo{journal}{Phys. Rev. Lett.}
  \textbf{\bibinfo{volume}{102}}, \bibinfo{pages}{167004}
  (\bibinfo{year}{2009}).

\bibitem[{\citenamefont{Coldea et~al.}(2009)\citenamefont{Coldea, Andrew,
  Analytis, McDonald, Bangura, Chu, Fisher, and Carrington}}]{Coldea09}
\bibinfo{author}{\bibfnamefont{A.~I.}~\bibnamefont{Coldea}},
  \bibinfo{author}{\bibfnamefont{C.~M.~J.}~\bibnamefont{Andrew}},
  \bibinfo{author}{\bibfnamefont{J.~G.}~\bibnamefont{Analytis}},
  \bibinfo{author}{\bibfnamefont{R.~D.}~\bibnamefont{McDonald}},
  \bibinfo{author}{\bibfnamefont{A.~F.}~\bibnamefont{Bangura}},
  \bibinfo{author}{\bibfnamefont{J.~H.} \bibnamefont{Chu}},
  \bibinfo{author}{\bibfnamefont{I.~R.}~\bibnamefont{Fisher}}, \bibnamefont{and}
  \bibinfo{author}{\bibfnamefont{A.}~\bibnamefont{Carrington}},
  \bibinfo{journal}{Phys. Rev. Lett.} \textbf{\bibinfo{volume}{103}},
  \bibinfo{pages}{026404} (\bibinfo{year}{2009}).

\bibitem[{\citenamefont{Tsubota et~al.}(2013)\citenamefont{Tsubota, Wakita,
  Nagao, Hiramatsu, Ishiga, Sunagawa, Ono, Kumigashira, Danura, Kudo
  et~al.}}]{Tsubota13}
\bibinfo{author}{\bibfnamefont{K.}~\bibnamefont{Tsubota}},
  \bibinfo{author}{\bibfnamefont{T.}~\bibnamefont{Wakita}},
  \bibinfo{author}{\bibfnamefont{H.}~\bibnamefont{Nagao}},
  \bibinfo{author}{\bibfnamefont{C.}~\bibnamefont{Hiramatsu}},
  \bibinfo{author}{\bibfnamefont{T.}~\bibnamefont{Ishiga}},
  \bibinfo{author}{\bibfnamefont{M.}~\bibnamefont{Sunagawa}},
  \bibinfo{author}{\bibfnamefont{K.}~\bibnamefont{Ono}},
  \bibinfo{author}{\bibfnamefont{H.}~\bibnamefont{Kumigashira}},
  \bibinfo{author}{\bibfnamefont{M.}~\bibnamefont{Danura}},
  \bibinfo{author}{\bibfnamefont{K.}~\bibnamefont{Kudo}}, \bibnamefont{et~al.},
  \bibinfo{journal}{Journal of the Physical Society of Japan}
  \textbf{\bibinfo{volume}{82}}, \bibinfo{pages}{073705}
  (\bibinfo{year}{2013}).

\bibitem[{\citenamefont{Dhaka et~al.}(2014)\citenamefont{Dhaka, Jiang, Ran,
  Bud'ko, Canfield, Harmon, Kaminski, Tomi\ifmmode~\acute{c}\else \'{c}\fi{},
  Valent\'\i, and Lee}}]{Dhaka14}
\bibinfo{author}{\bibfnamefont{R.~S.} \bibnamefont{Dhaka}},
  \bibinfo{author}{\bibfnamefont{R.}~\bibnamefont{Jiang}},
  \bibinfo{author}{\bibfnamefont{S.}~\bibnamefont{Ran}},
  \bibinfo{author}{\bibfnamefont{S.~L.} \bibnamefont{Bud'ko}},
  \bibinfo{author}{\bibfnamefont{P.~C.} \bibnamefont{Canfield}},
  \bibinfo{author}{\bibfnamefont{B.~N.} \bibnamefont{Harmon}},
  \bibinfo{author}{\bibfnamefont{A.}~\bibnamefont{Kaminski}},
  \bibinfo{author}{\bibfnamefont{M.}~\bibnamefont{Tomi\ifmmode~\acute{c}\else
  \'{c}\fi{}}}, \bibinfo{author}{\bibfnamefont{R.}~\bibnamefont{Valent\'i}},
  \bibnamefont{and} \bibinfo{author}{\bibfnamefont{Y.}~\bibnamefont{Lee}},
  \bibinfo{journal}{Phys. Rev. B} \textbf{\bibinfo{volume}{89}},
  \bibinfo{pages}{020511} (\bibinfo{year}{2014}).

\bibitem[{\citenamefont{Aichhorn et~al.}(2010)\citenamefont{Aichhorn, Biermann,
  Miyake, Georges, and Imada}}]{Aichhorn10}
\bibinfo{author}{\bibfnamefont{M.}~\bibnamefont{Aichhorn}},
  \bibinfo{author}{\bibfnamefont{S.}~\bibnamefont{Biermann}},
  \bibinfo{author}{\bibfnamefont{T.}~\bibnamefont{Miyake}},
  \bibinfo{author}{\bibfnamefont{A.}~\bibnamefont{Georges}}, \bibnamefont{and}
  \bibinfo{author}{\bibfnamefont{M.}~\bibnamefont{Imada}},
  \bibinfo{journal}{Phys. Rev. B} \textbf{\bibinfo{volume}{82}},
  \bibinfo{pages}{064504} (\bibinfo{year}{2010}).

\bibitem[{\citenamefont{Aichhorn et~al.}(2011)\citenamefont{Aichhorn,
  Pourovskii, and Georges}}]{Aichhorn11}
\bibinfo{author}{\bibfnamefont{M.}~\bibnamefont{Aichhorn}},
  \bibinfo{author}{\bibfnamefont{L.}~\bibnamefont{Pourovskii}},
  \bibnamefont{and} \bibinfo{author}{\bibfnamefont{A.}~\bibnamefont{Georges}},
  \bibinfo{journal}{Phys. Rev. B} \textbf{\bibinfo{volume}{84}},
  \bibinfo{pages}{054529} (\bibinfo{year}{2011}).

\bibitem[{\citenamefont{Yin et~al.}(2011)\citenamefont{Yin, Haule, and
  Kotliar}}]{Yin2011a}
\bibinfo{author}{\bibfnamefont{Z.~P.} \bibnamefont{Yin}},
  \bibinfo{author}{\bibfnamefont{K.}~\bibnamefont{Haule}}, \bibnamefont{and}
  \bibinfo{author}{\bibfnamefont{G.}~\bibnamefont{Kotliar}},
  \bibinfo{journal}{Nat. Mater.} \textbf{\bibinfo{volume}{10}},
  \bibinfo{pages}{932} (\bibinfo{year}{2011}).

\bibitem[{\citenamefont{Ferber et~al.}(2012{\natexlab{a}})\citenamefont{Ferber,
  Foyevtsova, Valenti, and Jeschke}}]{Ferber12b}
\bibinfo{author}{\bibfnamefont{J.}~\bibnamefont{Ferber}},
  \bibinfo{author}{\bibfnamefont{K.}~\bibnamefont{Foyevtsova}},
  \bibinfo{author}{\bibfnamefont{R.}~\bibnamefont{Valenti}}, \bibnamefont{and}
  \bibinfo{author}{\bibfnamefont{H.~O.} \bibnamefont{Jeschke}},
  \bibinfo{journal}{Phys. Rev. B} \textbf{\bibinfo{volume}{85}},
  \bibinfo{pages}{094505} (\bibinfo{year}{2012}{\natexlab{a}}).

\bibitem[{\citenamefont{Ferber et~al.}(2012{\natexlab{b}})\citenamefont{Ferber,
  Jeschke, and Valent\'{\i}}}]{Ferber2012a}
\bibinfo{author}{\bibfnamefont{J.}~\bibnamefont{Ferber}},
  \bibinfo{author}{\bibfnamefont{H.~O.}~\bibnamefont{Jeschke}}, \bibnamefont{and}
  \bibinfo{author}{\bibfnamefont{R.}~\bibnamefont{Valent\'i}},
  \bibinfo{journal}{Phys. Rev. Lett.} \textbf{\bibinfo{volume}{109}},
  \bibinfo{pages}{236403} (\bibinfo{year}{2012}{\natexlab{b}}).

\bibitem[{\citenamefont{Werner et~al.}(2012)\citenamefont{Werner, Casula,
  Miyake, Aryasetiawan, Millis, and Biermann}}]{Werner12}
\bibinfo{author}{\bibfnamefont{P.}~\bibnamefont{Werner}},
  \bibinfo{author}{\bibfnamefont{M.}~\bibnamefont{Casula}},
  \bibinfo{author}{\bibfnamefont{T.}~\bibnamefont{Miyake}},
  \bibinfo{author}{\bibfnamefont{F.}~\bibnamefont{Aryasetiawan}},
  \bibinfo{author}{\bibfnamefont{A.~J.} \bibnamefont{Millis}},
  \bibnamefont{and} \bibinfo{author}{\bibfnamefont{S.}~\bibnamefont{Biermann}},
  \bibinfo{journal}{Nat. Phys.} \textbf{\bibinfo{volume}{8}},
  \bibinfo{pages}{331} (\bibinfo{year}{2012}).

\bibitem[{\citenamefont{Blaha et~al.}(2001)\citenamefont{Blaha, Schwarz,
  Madsen, Kvasnicka, and Luitz}}]{Blaha01}
\bibinfo{author}{\bibfnamefont{P.}~\bibnamefont{Blaha}},
  \bibinfo{author}{\bibfnamefont{K.}~\bibnamefont{Schwarz}},
  \bibinfo{author}{\bibfnamefont{G.~K.~H.} \bibnamefont{Madsen}},
  \bibinfo{author}{\bibfnamefont{D.}~\bibnamefont{Kvasnicka}},
  \bibnamefont{and} \bibinfo{author}{\bibfnamefont{J.}~\bibnamefont{Luitz}},
  \emph{\bibinfo{title}{{An Augmented Plane Wave Plus Local Orbitals Program
  for Calculating Crystal Properties}}}, vol.~\bibinfo{volume}{1}
  (\bibinfo{year}{2001}).

\bibitem[{\citenamefont{Perdew et~al.}(1996)\citenamefont{Perdew, Burke, and
  Wang}}]{Perdew96}
\bibinfo{author}{\bibfnamefont{J.~P.} \bibnamefont{Perdew}},
  \bibinfo{author}{\bibfnamefont{K.}~\bibnamefont{Burke}}, \bibnamefont{and}
  \bibinfo{author}{\bibfnamefont{Y.}~\bibnamefont{Wang}},
  \bibinfo{journal}{Phys. Rev. B} \textbf{\bibinfo{volume}{54}},
  \bibinfo{pages}{16533} (\bibinfo{year}{1996}).

\bibitem[{\citenamefont{Aichhorn et~al.}(2009)\citenamefont{Aichhorn,
  Pourovskii, Vildosola, Ferrero, Parcollet, Miyake, Georges, and
  Biermann}}]{Aichhorn09}
\bibinfo{author}{\bibfnamefont{M.}~\bibnamefont{Aichhorn}},
  \bibinfo{author}{\bibfnamefont{L.}~\bibnamefont{Pourovskii}},
  \bibinfo{author}{\bibfnamefont{V.}~\bibnamefont{Vildosola}},
  \bibinfo{author}{\bibfnamefont{M.}~\bibnamefont{Ferrero}},
  \bibinfo{author}{\bibfnamefont{O.}~\bibnamefont{Parcollet}},
  \bibinfo{author}{\bibfnamefont{T.}~\bibnamefont{Miyake}},
  \bibinfo{author}{\bibfnamefont{A.}~\bibnamefont{Georges}}, \bibnamefont{and}
  \bibinfo{author}{\bibfnamefont{S.}~\bibnamefont{Biermann}},
  \bibinfo{journal}{Phys. Rev. B} \textbf{\bibinfo{volume}{80}},
  \bibinfo{pages}{085101} (\bibinfo{year}{2009}).

\bibitem{Ferber2014} J. Ferber, K. Foyevtsova, H. O. Jeschke, and
  R. Valent{\'\i}, Phys. Rev. B {\bf 89}, 205106 (2014).

\bibitem[{\citenamefont{Werner et~al.}(2006)\citenamefont{Werner, Comanac,
  de' Medici, Troyer, and Millis}}]{Werner06}
\bibinfo{author}{\bibfnamefont{P.}~\bibnamefont{Werner}},
  \bibinfo{author}{\bibfnamefont{A.}~\bibnamefont{Comanac}},
  \bibinfo{author}{\bibfnamefont{L.}~\bibnamefont{de' Medici}},
  \bibinfo{author}{\bibfnamefont{M.}~\bibnamefont{Troyer}}, \bibnamefont{and}
  \bibinfo{author}{\bibfnamefont{A.~J.}~\bibnamefont{Millis}},
  \bibinfo{journal}{Phys. Rev. Lett.} \textbf{\bibinfo{volume}{97}},
  \bibinfo{pages}{076405} (\bibinfo{year}{2006}).

\bibitem[{\citenamefont{Bauer et~al.}(2011)\citenamefont{Bauer, Carr, Evertz,
  Feiguin, Freire, Fuchs, Gamper, Gukelberger, Gull, Guertler et~al.}}]{ALPS11}
\bibinfo{author}{\bibfnamefont{B.}~\bibnamefont{Bauer}},
  \bibinfo{author}{\bibfnamefont{L.~D.} \bibnamefont{Carr}},
  \bibinfo{author}{\bibfnamefont{H.~G.} \bibnamefont{Evertz}},
  \bibinfo{author}{\bibfnamefont{A.}~\bibnamefont{Feiguin}},
  \bibinfo{author}{\bibfnamefont{J.}~\bibnamefont{Freire}},
  \bibinfo{author}{\bibfnamefont{S.}~\bibnamefont{Fuchs}},
  \bibinfo{author}{\bibfnamefont{L.}~\bibnamefont{Gamper}},
  \bibinfo{author}{\bibfnamefont{J.}~\bibnamefont{Gukelberger}},
  \bibinfo{author}{\bibfnamefont{E.}~\bibnamefont{Gull}},
  \bibinfo{author}{\bibfnamefont{S.}~\bibnamefont{Guertler}},
  \bibnamefont{et~al.}, \bibinfo{journal}{J. Stat. Mech. Theory Exp.}
  \textbf{\bibinfo{volume}{2011}}, \bibinfo{pages}{P05001}
  (\bibinfo{year}{2011}).

\bibitem[{\citenamefont{Gull et~al.}(2011)\citenamefont{Gull, Werner, Fuchs,
  Surer, Pruschke, and Troyer}}]{Gull11a}
\bibinfo{author}{\bibfnamefont{E.}~\bibnamefont{Gull}},
  \bibinfo{author}{\bibfnamefont{P.}~\bibnamefont{Werner}},
  \bibinfo{author}{\bibfnamefont{S.}~\bibnamefont{Fuchs}},
  \bibinfo{author}{\bibfnamefont{B.}~\bibnamefont{Surer}},
  \bibinfo{author}{\bibfnamefont{T.}~\bibnamefont{Pruschke}}, \bibnamefont{and}
  \bibinfo{author}{\bibfnamefont{M.}~\bibnamefont{Troyer}},
  \bibinfo{journal}{Comput. Phys. Commun.} \textbf{\bibinfo{volume}{182}},
  \bibinfo{pages}{1078} (\bibinfo{year}{2011}).

\bibitem[{\citenamefont{Anisimov et~al.}(1993)\citenamefont{Anisimov, Solovyev,
  and Korotin}}]{Anisimov93}
\bibinfo{author}{\bibfnamefont{V.~I.}~\bibnamefont{Anisimov}},
  \bibinfo{author}{\bibfnamefont{I.~V.}~\bibnamefont{Solovyev}},
  \bibinfo{author}{\bibfnamefont{M.~A.}~\bibnamefont{Korotin}}, 
  \bibinfo{author}{\bibfnamefont{M.~T.}~\bibnamefont{Czyzyk}},\bibnamefont{and}
  \bibinfo{author}{\bibfnamefont{G.~A.}~\bibnamefont{Sawatzky}},
  \bibinfo{journal}{Phys. Rev. B} \textbf{\bibinfo{volume}{48}},
  \bibinfo{pages}{16929}
  (\bibinfo{year}{1993}).

\bibitem[{\citenamefont{Dudarev and Botton}(1998)}]{Dudarev98}
  \bibinfo{author}{\bibfnamefont{S.~L.}~\bibnamefont{Dudarev}},
  \bibinfo{author}{\bibfnamefont{G.~A.}~\bibnamefont{Botton}},
  \bibinfo{author}{\bibfnamefont{S.~Y.}~\bibnamefont{Savrasov}},
  \bibinfo{author}{\bibfnamefont{C.~J.}~\bibnamefont{Humphreys}} \bibnamefont{and}
  \bibinfo{author}{\bibfnamefont{A.~P.}~\bibnamefont{Sutton}},
  \bibinfo{journal}{Phys. Rev. B} \textbf{\bibinfo{volume}{57}},
  \bibinfo{pages}{1505} (\bibinfo{year}{1998}).

\bibitem[{\citenamefont{Czy??yk and Sawatzky}(1994)}]{Czyzyk94}
\bibinfo{author}{\bibfnamefont{M.~T.}~\bibnamefont{Czyzyk}} \bibnamefont{and}
  \bibinfo{author}{\bibfnamefont{G.~A.}~\bibnamefont{Sawatzky}},
  \bibinfo{journal}{Phys. Rev. B} \textbf{\bibinfo{volume}{49}},
  \bibinfo{pages}{14211}
  (\bibinfo{year}{1994}).

\bibitem[{\citenamefont{Liechtenstein et~al.}(1995)\citenamefont{Liechtenstein,
  Anisimov, and Zaanen}}]{Liechtenstein95}
\bibinfo{author}{\bibfnamefont{A.~I.}~\bibnamefont{Liechtenstein}},
  \bibinfo{author}{\bibfnamefont{V.~I.}~\bibnamefont{Anisimov}}, \bibnamefont{and}
  \bibinfo{author}{\bibfnamefont{J.}~\bibnamefont{Zaanen}},
  \bibinfo{journal}{Phys. Rev. B} \textbf{\bibinfo{volume}{52}},
  \bibinfo{pages}{5467} (\bibinfo{year}{1995}).

\bibitem[]{milandiscuss}
\bibinfo{author}{\bibfnamefont{We have
  confirmed this trend by a tight-binding analysis of the
  bandstructure in the tetragonal and collapsed tetragonal phases
  (Milan Tomi\'c, private communication).}}

\bibitem[{\citenamefont{Haule and Kotliar}(2009)}]{Haule09}
\bibinfo{author}{\bibfnamefont{K.}~\bibnamefont{Haule}} \bibnamefont{and}
  \bibinfo{author}{\bibfnamefont{G.}~\bibnamefont{Kotliar}},
  \bibinfo{journal}{New Journal of Physics} \textbf{\bibinfo{volume}{11}},
  \bibinfo{pages}{025021} (\bibinfo{year}{2009}).

\bibitem[{\citenamefont{Gofryk et~al.}(2014)}]{Gofryk14}
\bibinfo{author}{\bibfnamefont{K.}~\bibnamefont{Gofryk}},
\bibinfo{author}{\bibfnamefont{B.}~\bibnamefont{Saparov}},
\bibinfo{author}{\bibfnamefont{T.}~\bibnamefont{Durakiewicz}},
\bibinfo{author}{\bibfnamefont{A.}~\bibnamefont{Chikina}},
\bibinfo{author}{\bibfnamefont{S.}~\bibnamefont{Danzenb\"acher}},
\bibinfo{author}{\bibfnamefont{D.~V.}~\bibnamefont{Vyalikh}},
\bibinfo{author}{\bibfnamefont{M.~J.}~\bibnamefont{Graf}} \bibnamefont{and}
\bibinfo{author}{\bibfnamefont{A.~S.}~\bibnamefont{Sefat}},
\bibinfo{journal}{Phys. Rev. Lett.} \textbf{\bibinfo{volume}{112}},
  \bibinfo{pages}{186401} (\bibinfo{year}{2014}).

\bibitem[{\citenamefont{Backes and Valenti}(2014)}]{k122}
\bibinfo{author}{\bibfnamefont{S.}~\bibnamefont{Backes}},
\bibinfo{author}{\bibfnamefont{D.}~\bibnamefont{Guterding}},
\bibinfo{author}{\bibfnamefont{H.~O.}~\bibnamefont{Jeschke}} \bibnamefont{and}
  \bibinfo{author}{\bibfnamefont{R.}~\bibnamefont{Valent\'i}},
  \bibinfo{journal}{arXiv:1403.6993} (\bibinfo{year}{2014}).

\bibitem[{\citenamefont{Mandal and Haule}(2014)}]{mandal14}
\bibinfo{author}{\bibfnamefont{S.}~\bibnamefont{Mandal}},
\bibinfo{author}{\bibfnamefont{R.~E.}~\bibnamefont{Cohen}} \bibnamefont{and}
  \bibinfo{author}{\bibfnamefont{K.}~\bibnamefont{Haule}},
  \bibinfo{journal}{arXiv:1407.4876} (\bibinfo{year}{2014}).

\end{thebibliography}

\end{document}